\documentclass[aps,prl,reprint,showpacs]{revtex4-1}

\usepackage{graphicx}
\usepackage{amsmath, amsthm, amssymb}
\usepackage{textcomp}

\usepackage{dcolumn}

\begin{document}

\title{Anderson wall and Bloch oscillations in molecular rotation}

\author{Johannes Flo\ss}
\author{Ilya Sh. Averbukh}
\affiliation{Department of Chemical Physics, Weizmann Institute of Science, 234 Herzl Street, Rehovot 76100, Israel}
\date{\today}

\begin{abstract}
We describe a universal behavior of linear molecules excited by a periodic train of short laser pulses under quantum resonance conditions.
In a rigid rotor the resonance causes an unlimited ballistic growth of the angular momentum.
We show that the centrifugal distortion of rotating molecules eventually halts the growth, by causing Anderson localization beyond a critical value of the angular momentum -- the Anderson wall.
Its position solely depends on the molecular rotational constants and lies in the range of a few tens of $\hbar$.
Below the wall, rotational excitation oscillates with the number of pulses due to a mechanism similar to Bloch oscillations in crystalline solids.
We suggest optical experiments capable of observing the rotational Anderson wall and Bloch oscillations at near-ambient conditions with the help of existing laser technology.
\end{abstract}



\maketitle



The periodically kicked rigid rotor is a standard model in nonlinear dynamics studies of many systems~\cite{Lichtenberg92,Casati06}.
The classical kicked rotor can exhibit truly chaotic motion leading to an unbounded diffusive growth of the angular momentum $J$.
In the quantum regime this chaotic motion is either suppressed by a mechanism similar to Anderson localization in disordered solids~\cite{anderson58} (as shown in~\cite{fishman82,*grempel84}), or the rotational excitation is enhanced due to the so-called quantum resonance~\cite{casati79,izrailev80}.
Although these fundamental quantum phenomena have been theoretically studied for several decades already, up to now there has not been a single experiment demonstrating dynamical Anderson localization in a real material rotating system.
Several known experiments on dynamical localization were done in a substitute system -- cold atoms interacting with a pulsed standing light wave -- imitating the dynamics of the planar kicked rotor~\cite{moore95,raizen99,klappauf98,amman98}.
An early proposal~\cite{bluemel86} on using microwave excited polar molecules as a testing ground for Anderson localization has never been realized (probably due to the complexity of the required field source). Recently~\cite{floss12,*floss12arxivshort,floss13}, we drew attention to the fact that current technology used for laser alignment of molecules offers tools for exploring the dynamics of the periodically kicked quantum rotor in a molecular system.
As a first step following our proposal~\cite{floss12},  quantum resonance excited by a few-pulse train was observed in molecular nitrogen~\cite{zhdanovich12a}.

In this Letter, we present  two qualitatively new rotational phenomena in linear molecules subject to moderately long resonant trains of laser pulses readily available nowadays.
We show that the resonant growth of the molecular angular momentum is suppressed  by centrifugal distortion, which causes Anderson localization beyond a critical value $J_{\mathrm{A}}$ of the angular momentum -- the Anderson wall.
Resonant rotational excitation that starts below $J_{\mathrm{A}}$ exhibits beats similar to Bloch oscillations of electrons in crystalline solids subject to a dc electric field~\cite{bloch1929,zener1934}.
The amplitude of the oscillations grows with the intensity of the laser pulses, but is restricted by the Anderson wall.
This wall and the oscillations are absent in the standard kicked rotor model or its atom optics implementation, but are universal features of kicked molecules.
We suggest several optical experiments on observing these phenomena.
Our results are of immediate importance for multiple current experiments employing resonant laser kicking for enhanced molecular alignment~\cite{cryan09}, isotope-selective excitation~\cite{zhdanovich12a,akagi12a}, and impulsive gas heating for Raman photoacoustics~\cite{schippers13} and controlling high power optical pulse propagation in atmosphere~\cite{zahedpour14}.

We consider  linear molecules interacting with short nonresonant linearly polarized laser pulses, as used in standard experiments on molecular alignment (for reviews see~\cite{fleischer12,ohshima10,stapelfeldt03}).
Here, the laser field affects the molecular rotation via Raman-type interaction~\cite{zon75,friedrich95,*friedrich95b}.
The electric field of the pulse induces anisotropic molecular polarization, interacts with it, and tends to align the molecular axis along the laser polarization direction.
An ultrashort laser pulse acts like a kick exciting molecular rotation, and the alignment is observed under field-free conditions after the pulse is over~\cite{ortigoso99,seideman99a,underwood05}.
The laser-molecule interaction with a pulse train is given by
\begin{equation}
V=-\frac{1}{4}\Delta\alpha \cos^2\theta \sum_{n=0}^{N-1}\mathcal{E}^2(t-n\tau)\,.
\label{eq.potential}
\end{equation}
Here, $\mathcal{E}(t)$ is the temporal envelope of a single laser pulse, $\Delta\alpha=\alpha_{\parallel}-\alpha_{\perp}$, where $\alpha_{\parallel}$ is the polarizability along the molecular axis and $\alpha_{\perp}$ is the polarizability perpendicular to it, $N$ is the number of pulses, $\tau$ is the period of the train, and $\theta$ is the angle between the laser polarization axis and the molecular axis.
We introduce the effective interaction strength $P=\Delta\alpha/(4\hbar)\int  \mathcal{E}^2(t) \mathrm{d}t $.
It reflects the typical change of the molecular angular momentum (in units of $\hbar$) induced by a single laser pulse.
Mind that the potential $V$ couples only states with $\Delta J=0,\pm2$ and $\Delta M_J=0$, where $J$ is the angular momentum and $M_J$ is its projection on the laser polarisation direction.

The rotational levels of a linear molecule are~\cite{herzbergbook}
\begin{equation}
E_J=BJ(J+1)-DJ^2(J+1)^2 \,,
\label{eq.energylevels}
\end{equation}
where $B$ and $D$ are the rotational and the centrifugal distortion constant, respectively.
For low-lying rotational states one may neglect the second term in Eq.~\eqref{eq.energylevels}, and the dynamics of a rigid molecular rotor is defined by a single parameter, the so-called rotational revival time, $t_{\mathrm{rev}}=\pi\hbar/B$.
Any rotational wave packet of a free rotor periodically reproduces itself after an integer multiple of the revival time.
This is the physical reason for the quantum resonance:
Short kicks separated in time by $t_{\mathrm{rev}}$ add constructively their actions, and a series of $N$ resonant kicks of strength $P$ is equivalent to a single kick of the strength $NP$.
As a result, the angular momentum of the molecule grows ballistically (linear) with $N$.
As the molecule rotates faster, the centrifugal force pulls the atoms apart, and due to the increase of the moment of inertia the molecule is gradually detuned from the initial quantum resonance, with consequences described below.

A good way to understand the dynamics of a periodically driven quantum system is by looking at its quasienergy states (Floquet states)~\cite{zeldovich67}, the eigenstates of a one-cycle (pulse-to-pulse) evolution operator.
The quasienergy eigenstate $|\chi_{\varepsilon} \rangle (t)$ reproduces itself after a one-period evolution up to a certain phase factor:
\begin{equation}\label{qes}
   |\chi_{\varepsilon} \rangle (t+\tau)=e^{-i\varepsilon \tau /\hbar }|\chi_{\varepsilon} \rangle (t) ,
\end{equation}
where $\varepsilon$ is the quasienergy eigenvalue.
It was shown in the past~\cite{fishman82} that finding quasienergy states for the kicked rotor problem can be mapped onto solving a stationary Schr\"{o}dinger equation for a one-dimensional tight-binding model, in which each site corresponds to a $J$ state and the energy of the site is proportional to
\begin{equation}
T(J)=\tan\left[\tau(\varepsilon - E_J )/(2\hbar)\right] .
\label{eq.ALcondition}
\end{equation}
For resonant kicking ($\tau=t_{\mathrm{rev}}$), this quantity becomes
\begin{equation}\label{res}
  T_{\mathrm{r}}(J)= \tan [\phi (J)] ; \, \phi (J)=\frac{\pi}{2}\left[\frac{\varepsilon}{B}+\frac{D}{B}J^2 (J+1)^2 \right].
\end{equation}
Here, we omitted the integer multiples of $\pi$ in $\phi (J)$.
For most molecules the ratio of $D/B$ is a small number (see Tab.~\ref{tab.ja}), so for small values of $J$ one may neglect the centrifugal distortion, and Eq.~\eqref{res} corresponds to a completely periodic lattice of identical sites.
The energy spectrum of such a system has a band structure related to an unlimited propagation along the lattice, as expected under the condition of quantum resonance.
For large $J$ the second term in the argument of the tangent in Eq.~\eqref{res} is important.
Using speculations similar to those of Ref.~\cite{fishman82}, one may argue that $T_{\mathrm{r}}(J)$ behaves as a pseudorandom function for large enough $J$.
Indeed, the difference of phases $\Delta \phi (J) =\phi (J+2) -\phi (J)$ for two neighboring sites becomes a large number $\sim 4 \pi (D/B) J^3$, rapidly growing with $J$.
Because of the $\pi$-periodicity of the tangent function, only $\Delta \phi (J) \mod (\pi )$ matters for the site-to-site variations of $T_{\mathrm{r}}(J)$.
The ratio of two spectroscopic constants $D/B$ is generally an irrational number.
Therefore, $\Delta \phi (J)$ is not simply related to $\pi$, which makes $T_{\mathrm{r}}(J)$ ``unpredictable'' and pseudorandom.
Under such conditions, we may expect the quasienergy states to become localized in $J$ space, similar to electronic states localized in disordered solids as predicted by Anderson in 1958~\cite{anderson58}.
The localization is likely to start when $\Delta \phi (J)$ reaches the value of $\sim \pi/2$, on the halfway between zero and the $\pi$-period of the tangent function in Eq.~\eqref{res}, i.e., at
\begin{equation}\label{wall}
 J_{\mathrm{A}} \sim \frac{1}{2}\sqrt[3]{\frac{B}{D}} \,.
\end{equation}
The value of $J_{\mathrm{A}}$ -- the ``Anderson wall'' separating delocalized and localized quasienergy states under resonant driving -- is a distinct attribute of every molecule.
In Tab.~\ref{tab.ja} we show its estimated value for several molecules.

\begin{table}
\begin{ruledtabular}
\begin{tabular}{cddd}
&\multicolumn{1}{c}{$t_{\mathrm{rev}}$ in ps} & \multicolumn{1}{c}{$D/B\cdot10^{6}$} & \multicolumn{1}{c}{estimated $J_{\mathrm{A}}$} \\
\hline
H$_2$&0.281&794&\sim5\\
N$_2$&8.383&2.90&\sim35\\
Cl$_2$&68.57&0.765&\sim55\\
ICl&146.4&0.354&\sim70\\
CO$_2$&42.74&0.343&\sim70\\
Br$_2$&203.5&0.255&\sim80\\
OCS&82.22&0.214&\sim80\\
I$_2$&447.0&0.107&\sim100\\
\end{tabular}
\end{ruledtabular}
\caption{\label{tab.ja}
Rotational revival time $t_{\mathrm{rev}}$, ratio $D/B$ of the centrifugal distortion constant and the rotational one, and estimated position $J_{\mathrm{A}}$ of the Anderson wall, for typical linear molecules.
All data is for the vibrational ground state.
Rotational constants are taken from~\cite{nist,val71,wells83}.
}
\end{table}

\begin{figure}
\includegraphics{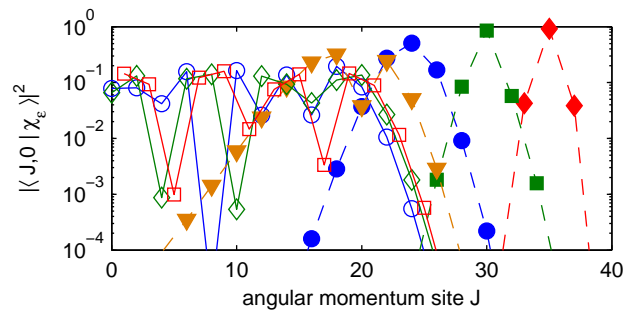}
\caption{\label{fig.statesCD}
Selected quasienergy states $|\chi_{\varepsilon}\rangle$ (projected on the angular momentum states $|J,0\rangle$) for resonantly kicked nonrigid $^{14}$N$_2$ molecules. Kick strength is $P=3$.
The quasienergy values for the presented states are (in the units of $\hbar/t_{\mathrm{rev}}$): 0.21 (triangles), 0.27 (full diamonds), 1.74 (full circles), 5.08 (empty circles), 5.44 (empty diamonds), 5.60 (empty squares), 6.27 (full squares).
}
\end{figure}

To test our arguments, we calculated the quasienergy states for  $^{14}$N$_2$ molecules kicked periodically at the rotational revival time $t_{\mathrm{rev}}=8.38~\mathrm{ps}$ by laser pulses of $P=3$.
This value corresponds to 50~fs long pulses (full width at half maximum of the intensity envelope) of peak intensities  of approximately  $40~\text{TW}/\text{cm}^2$, which is below the onset of ionization~\cite{cryan09,guo98}.
For the range of $J$-states considered here, such a pulse is shorter than the typical rotational periods, and it acts as a delta-pulse.
We have specially verified that the finite duration of the pulses has no qualitative influence on the results presented below.
The details of our numerical procedure can be found in~\cite{floss13}.
Figure~\ref{fig.statesCD} shows the absolute squares of the projection coefficients $|\langle J,0|\chi_{\varepsilon} \rangle |^2$ for several representative quasienergy states.
For low values of $J$, one can see extended states with an almost constant (in logarithmic scale) amplitude having a cutoff around $J_{\mathrm{R}}\sim 20$.
These extended states correspond to the initial ballistic growth of the angular momentum under the condition of quantum resonance.
For $J\geq J_{\mathrm{A}} \sim 35$, the quasienergy states are tightly localized around single angular momentum sites, in a good agreement with the above arguments on Anderson localization.
There is a transition region $J_{\mathrm{R}}<J<J_{\mathrm{A}}$ between the cutoff of the extended states and the Anderson wall, in which the quasienergy states span over a couple of momentum sites.
The cutoff position $J_{\mathrm{R}}\leq J_{\mathrm{A}}$ grows with $P$, while the position of the Anderson wall is $P$-independent.

The above structure of the quasienergy states  leads to some rather general conclusions  about the rotational dynamics of molecules subject to resonant kicking by laser pulses.
If a molecule is initially in a low lying rotational level, its wavefunction has  reasonable overlap with several extended quasienergy states, but practically no overlap with the localized states beyond the Anderson wall.
Therefore, the molecule has no chance to be excited above $J=J_{\mathrm{A}}$, even if it is driven at the condition of quantum resonance by a very long pulse train.
Moreover, as the initial rotational wavefunction overlaps only with a finite number of quasienergy states, the driven rotational dynamics is almost periodic, and the molecules should closely reconstruct their initial state many times in the course of the long enough pulse trains~\cite{hogg82}.
On the other hand, if the initial angular momentum is above $J_{\mathrm{A}}$, we expect the wavefunction to be stuck to its initial momentum due to Anderson localization.

\begin{figure}
\includegraphics{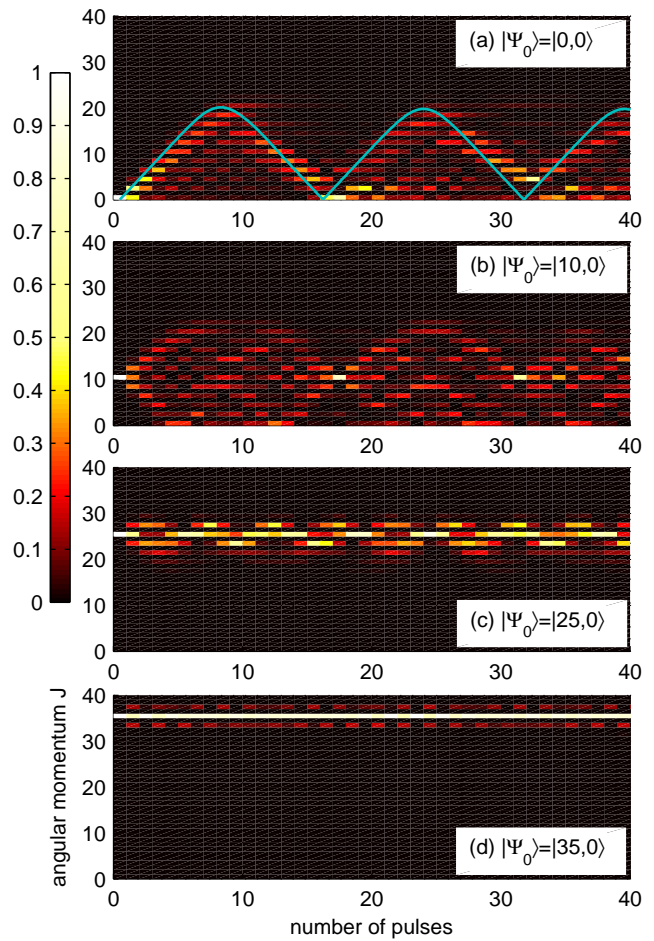}
\caption{\label{fig.distCD}
Simulated angular momentum distribution as a function of the number of pulses, for resonantly kicked $^{14}$N$_2$ molecules with initial state $|\Psi_0\rangle$.
Pulses are 50~fs long, with kick strength $P=3$.
The solid line in panel (a) is a solution of the semi-classical model~\eqref{eq.diff}.
}
\end{figure}

Figure~\ref{fig.distCD} demonstrates how these general statements reveal themselves in the above example of resonantly kicked $^{14}$N$_2$ molecules.
In the Figure, the simulated time-dependent angular distribution is shown as a function of the number of pulses.
The details of the numerical procedure can be found in our previous works~\cite{fleischer09,floss12b}.
For a molecule initially at rest [Fig.~\ref{fig.distCD}(a)], one can see first a ballistic growth of the angular momentum, where each pulse shifts the distribution by about 3 units of $\hbar$.
After about eight pulses, when the wave of excitation reaches $J_{\mathrm{R}}\approx 20$, it is reflected and ballistically propagates back towards $J=0$.
After about 16 pulses, the molecule (approximately) returns to the initial state, and the cycle starts again.
If the initial state is below $J_{\mathrm{R}}$ but above $J=0$ [Fig.~\ref{fig.distCD}~(b)], one can observe two streams of rotational excitation, directed upwards and downwards.
The physical reason for the double stream is that in half of the angular space the direction of the kick coincides with the initial rotational velocity of the molecule, whilst in the other half it is directed oppositely.
The two streams propagate until they reach the turning points $J=J_{\mathrm{R}}$ and $J=0$, and are reflected from there.
For an initial state lying in the interval $J_{\mathrm{R}}< J_0 < J_{\mathrm{A}}$ [Fig.~\ref{fig.distCD}(c)], the amplitude of the oscillations of the angular momentum distribution is strongly reduced.
Finally, if the initial state lies behind the Anderson wall [Fig.~\ref{fig.distCD}(d)], kicking with the periodicity of the rotational revival time hardly leads to any excitation or de-excitation.
Instead, the angular momentum distribution is stuck to the initial rotational state (Anderson localization).

To reveal the physical origin of the oscillations observed for $J\leq J_{\mathrm{R}}$, let us first consider the limit of relatively weak pulses, $P\ll 1$.
In this case, the potential of Eq.~\eqref{eq.potential} couples only the nearest neighbors in the $J$-lattice, and the change of the rotational excitation from pulse to pulse is small.
Therefore, our rotational system corresponds to a quantum particle moving continuously in a periodic $J$-lattice.
In the absence of centrifugal distortion, the lattice sites are identical, and the quasienergy states are characterized by a continuous dimensionless quasimomentum $\vec{k}$ describing propagation of the Bloch wavefunction along the  $J$-lattice~\cite{ashcroft76}.
The quasienergy spectrum of this ``particle'' has a band structure: $\varepsilon (k)= -\hbar P/(2t_{\mathrm{rev}})\cos(2k)$.
Due to the nonrigidity of the molecules, there is an additional ``potential'' $U(J)$, which is proportional to the single-cycle phase added to each rotational state by the centrifugal distortion correction of the  rotational levels Eq.~\eqref{eq.energylevels}: $U(J)=-\pi(D/B) J^2(J+1)^2$. Notice that this ``potential'' may be additionally controlled by a slight detuning of the train period from the resonant value.
In analogy to solid state physics~\cite{bloch1929,zener1934}, one can derive the following semiclassical equations of motion for the quasimomentum $k$ and the lattice coordinate $J$:
\begin{equation}
\frac{\mathrm{d}k}{\mathrm{d}n} =  -\frac{\mathrm{d} U(J)}{\mathrm{d} J} \,;\ \  \frac{\mathrm{d}J}{\mathrm{d}n} =\frac{t_{\mathrm{rev}}}{\hbar}\frac{\mathrm{d} \varepsilon (k)}{\mathrm{d} k}= P\sin(2k) \,.
\label{eq.diff}
\end{equation}
Here, the dimensionless time $n$ is a continuous analog of the number of pulses.
The first equation is simply Newton's second law, and the second one defines the group velocity of the Bloch waves.
As we will show elsewhere, the differential equations~\eqref{eq.diff} can also be obtained by the so called ``$\epsilon$-classics'' approach~\cite{wimberger03}, which requires only $J<J_A$, but not $P\ll1$.
Eqs.~\eqref{eq.diff} are similar to the ones describing Bloch oscillations of electrons in crystalline solids subject to a constant electric field~\cite{bloch1929,zener1934}, with the only difference that here the accelerating ``force'' $-\mathrm{d} U(J)/\mathrm{d} J$ is ``coordinate''-dependent.
Because of the negative sign of the ``potential'' $U(J)$, the corresponding unidirectional ``force'' initially leads to accelerated motion away from $J=0$.
However, on a longer time-scale, the solution for $J(n)$ oscillates due to Bragg reflection of the Bloch waves from the edge of the Brillouin zone of the $J$-lattice.
Eqs.~\eqref{eq.diff} may be solved analytically in quadratures, as will be shown elsewhere.
In Fig.~\ref{fig.distCD}(a) the solution for $J(n)$ in the domain $J\geq0$ is plotted as a solid line.
As the initial quasimomentum we chose $k(0)=\pi/4$, as it corresponds to the initial growth rate $\mathrm{d}J/\mathrm{d}n=P$ of the angular momentum, which is the typical change of $J$ by a single pulse (see above).
One can see a very good agreement between the exact quantum mechanical solution and the semiclassical model of Eq.~\eqref{eq.diff}.

\begin{figure}
\includegraphics{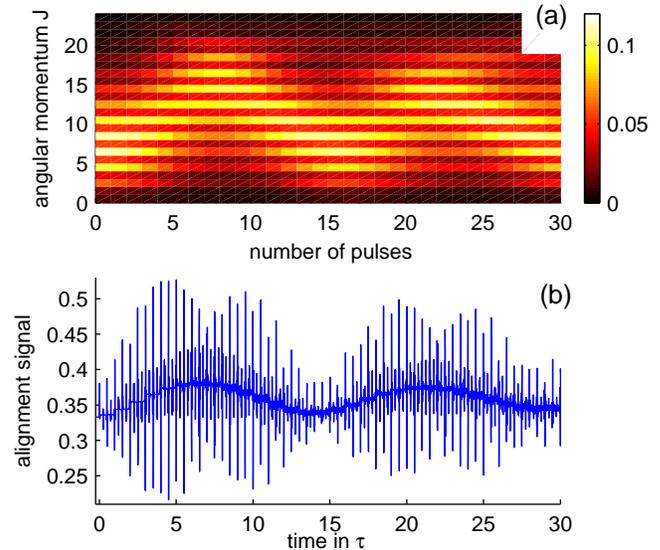}
\caption{\label{fig.distalCD}
Simulated (a) angular momentum distribution and (b) alignment signal $\langle \cos^2\theta \rangle$ as a function of the number of pulses, for resonantly kicked $^{14}$N$_2$ molecules.
Pulses are 50~fs long with kick strength $P=3$.
The initial temperature is $T=298~\mathrm{K}$.
The different intensity of even and odd states is due to nuclear spin statistics~\cite{herzbergbook}.
}
\end{figure}

The non-stationary rotational distribution can be routinely measured with time-resolved coherent Raman scattering spectroscopy, especially utilizing schemes in which full spectral information is retrieved with the use of a single femtosecond pulse~\cite{frostig11}.
Moreover, the evolution of the angular momentum distribution is  exhibited in the time-dependent alignment signal, defined as the expectation value $\langle \cos^2\theta \rangle (t)$.
The deviation of this quantity from the isotropic value  $1/3$  determines the laser induced anisotropy of the gas refraction index (birefringence), which can be measured by optical methods~\cite{renard03,fleischer12,ohshima10,stapelfeldt03}.
In Fig.~\ref{fig.distalCD}, the angular momentum distribution and the alignment signal for resonantly kicked nitrogen molecules at room temperature are shown.
The finite temperature is included in the simulation by calculating the signal for different initial states and summing the results, weighted by the corresponding Boltzmann factors (see~\cite{fleischer09,floss12b}).
No collisions are assumed to happen during the pulse train duration.
Remarkably, the saw-like pattern seen in Fig.~\ref{fig.distCD} survives quite well for the incoherent thermal initial state.
Regarding the alignment signal, one can see low-frequency oscillations of the baseline of the alignment signal (time-averaged alignment).
They can be attributed to the Bloch oscillations of the angular momentum distribution.
The conditions used for our simulations shown in Fig.~\ref{fig.distalCD} are very close to the conditions of the experiment described in~\cite{cryan09}, where molecular alignment induced by a periodic train of laser pulses was measured.
This experiment used 50~fs 800~nm pulses at 1~kHz repetition rate with peak intensities of 36~TW/cm$^2$, separated in time by approximately 8.4~ps, interacting with molecular nitrogen at STP conditions.
By comparing our Fig.~\ref{fig.distalCD}~(b) with Fig.~2 from Ref.~\cite{cryan09}, we conclude that Cryan~\textit{et al.} had actually reached the reflection point $J_{\mathrm{R}}$, but did not observe the Bloch oscillation itself, since their number of pulses was limited to eight.
A decisive experiment to monitor a full Bloch oscillation cycle requires about 16 pulses.
The optical scheme generating such a long periodic pulse train was already successfully demonstrated in~\cite{siders98}.
The total length of such a pulse train is about 125~ps, which is comparable to a single collision time at STP conditions.
Potential detrimental decoherence effects can be avoided by a slight reduction of the pressure used in~\cite{cryan09}.
An alternative route to detect the Bloch oscillations of the angular momentum distribution is to measure the total rotational energy absorbed by the molecules as a function of the number of pulses.
This can be done with the help of the photoacoustic approach~\cite{kartashov06,schippers11}, which measures the heat dumped into the gas as a result of rotational relaxation.
To demonstrate complete localization above the Anderson wall, one may use the optical centrifuge technique~\cite{karczmarek99,villeneuve00,yuan11,korobenko13} to prepare molecules in the high-$J$ initial states.

Concluding, we described two interrelated  phenomena -- Anderson wall and Bloch oscillations -- for linear molecules that are excited by a periodic train of short laser pulses under the condition of quantum resonance.
These effects are rotational counterparts of quantum localization phenomena in solid state physics, and can be observed in common molecules, like N$_2$, CO$_2$, OCS, etc., using existing schemes for generating laser pulse trains and a variety of available detection methods.
If successful, such experiments may result in the first observation of dynamical Anderson localization in a material kicked rotor.
As shown in our paper, laser experiments on rotating molecules provide a new testing ground for quantum localization effects, in addition to semiconductor super-lattices~\cite{feldmann92}, cold atoms in optical lattices (for a review, see~\cite{ibloch2005,kolovsky04}), and coupled photonic structures~\cite{morandotti99,sapienza03,christodoulides03,schwartz07,segev13}.
Most remarkably, these effects can be observed at near-ambient conditions.
Finally, our predictions have immediate practical implications for numerous current experimental schemes using resonant laser kicking for enhanced molecular alignment~\cite{cryan09}, isotope-selective excitation~\cite{zhdanovich12a,akagi12a}, and impulsive gas heating for Raman photoacoustics~\cite{schippers13} and controlling high power optical pulse propagation in atmosphere~\cite{zahedpour14}.

We appreciate multiple discussions related to the problem with Phil Bucksbaum, Shmuel Fishman, Andrei Kamalov, Yehiam Prior, Yaron Silberberg, and Uzy Smilansky.
We thank Victor L'vov for a discussion and critical reading of the manuscript.
I.~A.~acknowledges support as the Patricia Elman Bildner Professorial Chair.
This research was made possible in part by the historic generosity of the Harold Perlman Family.


%

\end{document}